\documentclass[a4paper,11t]{article}
\usepackage[top=2.54cm,bottom=2.54cm, left=2cm, right=2cm]{geometry}
\usepackage{graphicx} 
\usepackage{booktabs}
\usepackage{tabu}
\usepackage{amsmath,amssymb}
\usepackage{amsfonts}
\usepackage{float}
\usepackage{color}
\usepackage{changepage}
\usepackage{dirtytalk}
\usepackage{alphabeta}
\usepackage{pdflscape}
\usepackage{multirow}
\usepackage{bm}
\usepackage{enumitem}
\usepackage{enumerate}
\usepackage[colorlinks=true, allcolors=blue]{hyperref}
\usepackage[export]{adjustbox}
\usepackage[sort,comma]{natbib}
\usepackage{centernot}
\newcommand{\CI}{\mathrel{\perp\mspace{-10mu}\perp}}
\newcommand{\nCI}{\centernot{\CI}}

\title{On independence testing using the (partial) distance correlation}
\author{
Kontemeniotis Nikolaos\\
\href{mailto:kontemeniotisn@gmail.com}{kontemeniotisn@gmail.com} 
\and
Vargiakakis Rafail\\
\href{mailto:rafailvargiakakis@gmail.com}{rafailvargiakakis@gmail.com} 
\and
Tsagris Michail\\
\href{mailto:mtsagris@uoc.gr}{mtsagris@uoc.gr} \\
 \\
Department of Economics, University of Crete, Gallos Campus, Rethimnon, Greece
}

\begin{document}

\maketitle

\begin{center}
\textbf{Abstract}
\end{center}
Distance correlation is a measure of dependence between two paired random vectors or matrices of arbitrary, not necessarily equal, dimensions. Unlike Pearson correlation, the population distance correlation coefficient is zero if and only if the random vectors are independent. Thus, distance correlation measures both linear and non-linear association between two univariate and or multivariate random variables. Partial distance correlation expands to the case of conditional independence. To test for (conditional) independence, the p-value may be computed either via permutations or asymptotically via the $\chi^2$ distribution. In this paper we perform an intra-comparison of both approaches for (conditional) independence and an inter-comparison to the classical Pearson correlation where for the latter we compute the asymptotic p-value. The results are rather surprising, especially for the case of conditional independence.     

\section{Introduction}
In many real-world datasets, relationships between variables are not strictly linear. A non-linear association refers to a relationship where the change in one variable does not result in a constant change in another. Such associations are common in fields like biology, economics, and environmental science \citep{harrell2015}. 

The most popular association measures still remains the Pearson correlation coefficient despite being unable to capture non-linear associations between pairs of variables. Despite this drawback, the ease of computation and understanding of this measure partially explain its wide applicability across most disciplines. 

Distance correlation \citep{Szekely2007MeasuringAT} is a relatively new measure of dependence between two random variables that is sensitive to any form of association, including non-linear relationships. It has been praised as "a correlation for the 21st century" \cite{praise}, which further confirms its reliability in detecting any kind of relationship between two variables. Unlike the traditional Pearson correlation coefficient, which only captures linear relationships, distance correlation can detect a wide range of dependencies, and it is only equal to zero only if the two random variables are independent. 

The distance correlation has been extensively studied under the variable selection framework, see for instance \cite{li2012,zhong2015,yenigun2015,lu2020,tan2022}. For applications in the time series literature see \cite{davis2018} and the review of \cite{edelmann2019}.

Pearson correlation examines the linear association between two variables, which if assumed to be normal can extend the concept of linear independence to independence in general. Distance correlation on the other hand is used to evaluate the independence between two variables, that is linear or non-linear. 

Testing for linear independence using Pearson correlation is straightforward and relies upon the assumption of asymptotic normality. When it comes to distance correlation, to test for independence, computation of the permutation-based p-value is the classical approach. Relying upon distance metrics, distance correlation can be very computationally expensive for permutation based hypothesis testing. This is because given a vector $\bm{x}$ containing $n$ elements, the number of pairwise distances is equal to $n(n-1)/2$. To alleviate the computational cost associated with permutations, \cite{Shen02012022} proposed an asymptotic $\chi^2$ test that requires considerably less computations compared to the permutation test. 

In this paper we examine and compare the type I and power rates of both the Pearson and distance correlation, yielding both an inter- and an intra-comparison. Comparison between Pearson and distance correlation, and comparison between the asymptotic and the bootstrap based computation of the p-value of the distance correlation. We examine the performance of Pearson correlation in assessing nonlinear associations, but we also examine the performance of the distance correlation in order to decide whether or not the results given by the permutation test are worth the extra computational cost. The results of this investigation are expected to elucidate the strengths and weaknesses of permutation-based and asymptotic methods in distance correlation testing. By exploring their efficacy across varied conditions, this study aims to furnish researchers and practitioners with actionable recommendations for selecting the most suitable significance testing method for their data.

The paper is structured as follows: Section \ref{sec:definitions} briefly lists the necessary information regarding distance correlation, while Section \ref{sec:experimental} elaborates on the experimental design, including data generation, and the p-value computation. Section \ref{sec:conclusions} concludes with a summary of the study’s primary contributions, discusses the implications of these findings and proposes avenues for future research.

\section{Definitions} \label{sec:definitions}

\subsection{Pearson correlation}
The Pearson product-moment correlation coefficient (or Pearson correlation coefficient) is a measure of the strength of a linear association between two variables and is denoted by r or with the Greek letter $\rho$ when we refer to the population correlation coefficient which is a widely used tool until today to assess correlation. The values r takes are ranging from -1 to 1. A value of -1 indicates a perfect negative linear relationship, where an increase in one variable perfectly predicts a proportional decrease in the other. A value of 1 signifies a perfect positive linear relationship, with an increase in one variable predicting a proportional increase in the other. A value of 0 denotes no linear relationship, meaning changes in one variable do not predict changes in the other. The correlation coefficient between $x$ and $y$ is defined by:
\begin{eqnarray*}  
r_{xy} = \frac{\sum_{i=1}^{n}(x_i - \bar{x})(y_i - \bar{y})}{{\sqrt{\sum_{i=1}^{n}(x_i - \bar{x})^2}}{\sqrt{\sum_{i=1}^{n}(y_i - \bar{y})^2}}},
\end{eqnarray*}
Where $ \bar{x}$ and $\bar{y}$  are the sample means of $x_i$'s and $ y_i$'s, respectively.

The partial correlation coefficient, denoted as $r_{X,Y|Z}$, measures the relationship between two random variables $X$ and $Y$ excluding the effects of a control variable $Z$. It is used to test conditional independence and is defined as the linear correlation between the residuals from the linear regression of $ X$ on $Z$ and the linear regression of $Y$ on $Z$. The formula for the partial correlation coefficient $r_{X,Y|Z}$ is given by:
\begin{eqnarray*}   
r_{X,Y|Z} = 
\begin{cases} 
\frac{R_{X,Y} - R_{X,Z} R_{Y,Z}}{\sqrt{(1 - R_{X,Z}^2)(1 - R_{Y,Z}^2)}} & \text{if } |Z| = 1 \\ 
-\frac{A_{1,2}}{\sqrt{A_{1,1} A_{2,2}}} & \text{if } |Z| > 1 
\end{cases},
\end{eqnarray*}
Where  $R_{X,Y}$ is the correlation between variables $X$ and $Y$, $R_{X,Z}$ and $R_{Y,Z}$ denote the correlations between $X$ and $Z$, and $Y$ and $Z$, respectively. Let $A = R^{-1}_{X,Y,Z}$, where $A$ represents the correlation sub-matrix of variables $X, Y, Z$, and $A_{i,j}$ denotes the element in the $i$-th row and $j$-th column of matrix $A$.

\subsection{Distance correlation}
The mathematical formulation of the distance correlation measure relies on distance covariance which is calculated using the Euclidean distance matrices of the two random variables in question. For this purpose, let $\bm{X} \in R^p$ and $\bm{Y} \in R^q$ be two sets of data and $A,B$ be their Euclidean distance matrices respectively, where $\alpha_{ij}=||\bm{X}_i-\bm{X}_j||$ and $b_{ij}=||\bm{Y}_i-\bm{Y}_j||$. Using these elements we can next define the two corresponding double centered distance matrices:
\begin{eqnarray*}
    \hat{\bm{A}}_{ij} &=& a_{ij} - \bar{a}_{i.} - \bar{a}_{.j} + \bar{a}_{..}\\
    \hat{\bm{B}}_{ij} &=& b_{ij} - \bar{b}_{i.} - \bar{b}_{.j} + \bar{b}_{..}
\end{eqnarray*}
where
\begin{eqnarray*}   
\bar{a}_{i.}=\frac{1}{n}\sum_{j=1}^n \alpha_{ij},\hspace{0.3cm}
\bar{a}_{.j}=\frac{1}{n}\sum_{i=1}^n \alpha_{ij},\hspace{0.3cm}
\bar{a}_{..}=\frac{1}{n^2} \sum_{i,j=1}^{n}\alpha_{ij}.
\end{eqnarray*}

The entries of $B$ are defined similarly, and $n$ denotes the sample size. We use these matrices to construct the sample distance covariance, and variance measures \cite{Szekely2007MeasuringAT}:
\begin{eqnarray*}
dV_n^2(\bm{X},\bm{Y}) &=& \frac{1}{n^2}\sum_{i,j=1}^n \hat{A}_{ij}\hat{B}_{ij}\\
dV_n^2(\bm{X}) &=& V_n^2(\bm{X},\bm{X})=\frac{1}{n^2}\sum_{i,j=1}^n \hat{A}_{ij} \\
\end{eqnarray*}

Finally, we present an alternative formula as well as an unbiased estimator for the distance variance and covariance given by \cite{szekely2023energy}:
\begin{eqnarray*}
dV_n^2(\bm{X},\bm{Y}) &=& \frac{\sum_{i\ne j}^n \alpha_{ij}b_{ij}}{n^2}-2\frac{\sum_{i=1}^n\alpha_{i.}b_{i.}}{n^3}+\frac{\alpha_{..}b_{..}}{n^4} \\
dV_n^2(\bm{X}) &=& \frac{\sum_{i\ne j}^n \alpha_{ij}^2}{n^2}-2\frac{\sum_{i=1}^n\alpha_{i.}^2}{n^3}+\frac{\alpha_{..}^2}{n^4}.
\end{eqnarray*}

By altering the denominators we obtain the bias-corrected distance covariance. This is needed because the traditional distance correlation has a small-sample bias: its expected value does not approach zero under independence for small sample sizes. This means it tends to overestimate correlation when the true correlation is zero. To address this, we use the unbiased estimators of distance covariance, variance and correlation are
\begin{eqnarray*}
dV_n^{*2}(\bm{X},\bm{Y}) &=& \frac{\sum_{i\ne j}^n \alpha_{ij}b_{ij}}{n(n-3)}-2\frac{\sum_{i=1}^n\alpha_{i.}b_{i.}}{n(n-2)(n-3)}+\frac{\alpha_{..}b_{..}}{n(n-1)(n-2)(n-3)} \\
dV_n^{*2}(\bm{X}) &=& \frac{\sum_{i\ne j}^n \alpha_{ij}^2}{n(n-3)}-2\frac{\sum_{i=1}^n\alpha_{i.}^2}{n(n-2)(n-3)}+\frac{\alpha_{..}^2}{n(n-1)(n-2)(n-3)} \\
dR^{*2}_n(\bm{X},\bm{Y}) &=& \frac{dV^{*2}_n(\bm{X},\bm{Y})}{\sqrt{dV^{*2}_n(\bm{X}) dV_n^2(\bm{Y})}} \\
\end{eqnarray*}

We can also then define the partial distance correlation coefficient, which is used to measure the relationship between two variables while controlling for the influence of one or more additional variables. This helps isolate the true relationship between our $X$ and $Y$, by removing the effects of confounding factors. Given a random variable $Z$, the partial distance correlation of $X,Y$ is defined as \citep{szekely2014}
\begin{eqnarray*}
pdR^{*2}_n(\bm{X},\bm{Y}|\bm{Z}) = \frac{dR^{*2}_n(\bm{X},\bm{Y})-dR^{*2}_n(\bm{X},\bm{Z})dR_n^2(\bm{Y},\bm{Z})}{\sqrt{1-dR^{*2}_n(\bm{X},\bm{Z})^2}\sqrt{1-dR^{*2}_n(\bm{Y},\bm{Z})^2}}.
\end{eqnarray*}

A direct implementation of the distance correlation results in an \( O(n^2) \) computational complexity. To address this inefficiency, the paper by \citep{Huo2016} proposed a fast algorithm that reduces the computational complexity to \( O(n \log n) \), on average, When $p=q=1$, i.e. $X$ and $Y$ are vectors.

\subsection{Independence testing}
We addressed two scenarios of independence. First, marginal independence between two variables $X$ and $Y$ and conditional independence given a third variable $Z$ using Pearson and distance correlation. For distance correlation, we describe two methods to compute p-values: a permutation test and an asymptotic chi-square test. For Pearson correlation, we focus on the asymptotic approach. For given pairs of variables $(X, Y) \in \mathbb{R}$, assume they are independently and identically distributed as $F_{X,Y}$. The statistical hypothesis for testing independence is formulated as:
\begin{align*}
    H_0 &: F_{X,Y} = F_X F_Y, \\
    H_1 &: F_{X,Y} \neq F_X F_Y.
\end{align*}

We can do the same for conditional independence of observations $(X, Y, Z) \in \mathbb{R}$, assume they are independently and identically distributed as $F_{X,Y,Z}$. The statistical hypothesis for testing conditional independence of $X$ and $Y$ given $Z$ can be formulated as:
\begin{eqnarray*}    
\begin{aligned}
H_0 &: F_{X,Y \,\mid\, Z} \;=\; F_{X \,\mid\, Z}\, F_{Y \,\mid\, Z},\\
H_1 &: F_{X,Y \,\mid\, Z} \;\neq\; F_{X \,\mid\, Z}\, F_{Y \,\mid\, Z}.
\end{aligned}
\end{eqnarray*}

\subsubsection{Linear independence testing for the Pearson (partial) correlation}
For Pearson correlation to test the hypothesis that there is true correlation, $\rho$, equal to some pre-specified value $\rho_0$, we write:
\begin{eqnarray*}
\begin{aligned}
H_0 &: \rho = \rho_0,\\
H_1 &: \rho \neq \rho_0.
\end{aligned}
\end{eqnarray*}

We use Fisher's transformation:
\begin{eqnarray*}
F(\alpha) \;=\; \frac{1}{2}\,\ln\!\Bigl(\frac{1+\alpha}{1-\alpha}\Bigr).
\end{eqnarray*}

For the Pearson correlation coefficient $r$, the test statistic is given by:
\begin{eqnarray*}
T_0 = \left[F(r) - F(\rho_0)\right] \sqrt{n - 3}.
\end{eqnarray*}

Continuing to test for partial correlation, we can formulate the hypothesis:
\begin{eqnarray*}
\begin{aligned}
H_0 : \rho_{XY \mid \bm{Z}} = \rho_{0,XY \mid \bm{Z}},
\\
H_1 : \rho_{XY \mid \bm{Z}} \neq \rho_{0,XY \mid \bm{Z}}.
\end{aligned}
\end{eqnarray*}

Again we can employ Fisher’s transformation$(F(\alpha) \;=\; \frac{1}{2}\,\ln\!\Bigl(\frac{1+\alpha}{1-\alpha}\Bigr))$ as demonstrated above and the corresponding test statistic is then given by:
\begin{eqnarray*}
T_0 = \left[F(r_{XY \mid \bm{Z}}) - F(\rho_{0,XY \mid \bm{Z}}\right]\sqrt{n - 3 - |\bm{Z}|}.
\end{eqnarray*}
where $|\bm{Z}|$ denotes the number of control variables. Under $H_0$ both $T_0s$ are asymptotically distributed as $N(0, 1)$, but for small sample sizes the $t$ distribution with $n-3-|\bm{Z}|$ degrees of freedom is preferred. The inference proceeds similarly as simple Pearson correlation, allowing us to test whether the partial correlation differs significantly from the specified value.

\subsubsection{Independence testing for the (partial) distance correlation}
Currently there are two alternatives to obtain the p-value of independence using the (partial) distance correlation. The first suggested method is a permutation-based approach, which involves the computation of the $p$-value using random permutations. The second method is an asymptotic approximation from the $\chi^2$ distribution that replaces permutations. The test statistic is computed as $S = n \times dR^{*2}_n(X,Y)^2 + 1$ and $ S = n \times  pdR^{*2}_n(X,Y|Z) + 1$ for the distance and partial distance correlation, respectively, and the asymptotic p-value is derived as the upper tail of the $\chi^2$ distribution.

\section{Experimental Design} \label{sec:experimental}
We assessed the performance of the distance correlation test and Pearson correlation test under the null hypothesis of independence across various sample sizes and distributions. We generated independent samples from distinct sets of distributions: five different distributions for $X$ and three distinct distributions for $Y$. The specific distributions used to generate these independent samples are as follows: Beta distribution ($\mathbf{Be}$), skew-normal distribution ($\mathbf{SN}$), Cauchy Distribution ($\mathbf{Cau}$), Gamma distribution ($\mathbf{Ga}$), Von Mises distribution ($\mathbf{vM}$), mixtures of two normal distributions ($\mathbf{ M_2N}$), mixtures of three normal distributions ($\mathbf{M_3N}$) and mixtures of two skew-t distributions ($\mathbf{M_2St}$). 

These result in total of 15 unique scenarios. For each scenario, the experiment was replicated 1,000 times to ensure the robustness of our findings. We calculated the percentage of false non-rejections of the null hypothesis and varied the sample size ($n = (~50, ~100, ~200, ~500, ~1000, ~2000, ~5000, ~10000~)$) while performing two versions of the distance correlation test. 

The objective was to assess the nominal Type I error ($\alpha=5\%$) under the null hypothesis of (conditional) independence. We performed 1,000 replications for each scenario and sample size combination as in the previous case. We computed permutation and asymptotic p-values for the partial distance correlation by using $R_{perm} = 500$ permutations, and denote them by "\textit{P}" and "\textit{A}", respectively. The Pearson correlation coefficient is denoted by "\textit{Pear}". 

Below, we define the empirical Type I error rate for either method as follows:
\begin{eqnarray*}
\hat{\alpha}_j(n) = \frac{1}{1000} \sum_{i=1}^{1000} I\left(p_{ij} < 0.05\right),
\end{eqnarray*}
where $I(\cdot)$ is an indicator function that equals 1 if the condition is true and 0 otherwise, and $p_{i,j}$ is the $i$-th p-value (permutation or asymptotic) under the $j$-th scenario. Under nominal Type I error control and the null hypothesis of independence, we expect \(\hat{\alpha}_{j}(n) \approx 0.05\).

All experiments were conducted on a laptop with Linux Mint 22.1 x86$\_$64 installed, AMD Ryzen 5 5600H processor and 16GB RAM. To compute the permutation-based p-value for the distance correlation test the function \texttt{dcor.test()} from the \textit{R} package \textsf{dcov} \citep{dcov2020} was used, while for the  partial distance correlation the function \texttt{pdcor.test()} from the \textit{R} package \textsf{pdcor} \citep{pdcor2025} was used.

\subsection{Case 1: Independence}
This is a simple example of independence using the 15 aforementioned pairs. The results on Table \ref{Simple Dcor} show that Pearson correlation seems to perform pretty accurately in terms of attaining the correct size of the test, and this is most evident for the larger sample sizes. The permutation-based distance correlation test seems to perform equally well, whereas the asymptotic test is conservative in two cases, namely the pairs of Cauchy \& Von Mises distributions and mixtures of two skew-t distributions, $\mathbf{Cau}-\mathbf{M_2St}$ and 
$\mathbf{vM}-\mathbf{M_2St}$, respectively.

\begin{table}[h!]
\caption{Estimated type I error for Case 1: independence.}
\centering
\begin{tabu} to \linewidth {l|>{\raggedright}p{0.8cm}>{\raggedleft}p{0.8cm}>{\raggedleft}p{0.8cm}>{\raggedleft}p{0.8cm}>{\raggedleft}p{0.8cm}>{\raggedleft}p{0.8cm}>{\raggedleft}p{0.8cm}>{\raggedleft}p{0.8cm}>{\raggedleft}p{0.75cm}}
\toprule
\toprule
  \textbf{Distributions}& \textbf{N} & \textbf{50} & \textbf{100} & \textbf{200} & \textbf{500} & \textbf{1000} & \textbf{2000} & \textbf{5000} & \textbf{10000}\\
\midrule
\midrule
\multirow{3}{*}{$\mathbf{Be - M_2N}$} & P & 0.049 & 0.060 & 0.045 & 0.039 & 0.051 & 0.054 & 0.053 & 0.045\\
& A & 0.053 & 0.060 & 0.043 & 0.036 & 0.047 & 0.049 & 0.047 & 0.040\\
& Pear &0.046 & 0.038 & 0.050 & 0.040 & 0.047 & 0.053 & 0.057 & 0.066\\
\hline
\multirow{3}{*}{$\mathbf{Be - M_3N}$} & P & 0.055 & 0.046 & 0.050 & 0.060 & 0.049 & 0.063 & 0.051 & 0.047\\
& A & 0.053 & 0.047 & 0.050 & 0.060 & 0.041 & 0.054 & 0.051 & 0.044\\
& Pear &0.049 & 0.050 & 0.047 & 0.038 & 0.052 & 0.053 & 0.064 & 0.050
\\
\hline
\multirow{3}{*}{$\mathbf{Be - M_2St}$} & P & 0.055 & 0.050 & 0.045 & 0.064 & 0.040 & 0.051 & 0.055 & 0.048\\
& A & 0.051 & 0.050 & 0.047 & 0.058 & 0.031 & 0.046 & 0.048 & 0.039\\
& Pear &0.042 & 0.048 & 0.051 & 0.053 & 0.057 & 0.050 & 0.049 & 0.045\\
\hline
\multirow{3}{*}{$\mathbf{Cau - M_2N}$} & P & 0.050 & 0.056 & 0.045 & 0.044 & 0.046 & 0.047 & 0.047 & 0.046\\
& A & 0.049 & 0.053 & 0.037 & 0.036 & 0.036 & 0.042 & 0.039 & 0.037\\
& Pear &0.048 & 0.049 & 0.065 & 0.050 & 0.058 & 0.041 & 0.062 & 0.048\\
\hline
\multirow{3}{*}{$\mathbf{Cau - M_3N}$} & P & 0.060 & 0.052 & 0.054 & 0.052 & 0.059 & 0.051 & 0.044 & 0.060\\
& A & 0.057 & 0.043 & 0.047 & 0.045 & 0.047 & 0.040 & 0.039 & 0.051\\
& Pear &0.049 & 0.045 & 0.044 & 0.066 & 0.055 & 0.039 & 0.059 & 0.047\\
\hline
\multirow{3}{*}{$\mathbf{Cau - M_2St}$} & P & 0.060 & 0.064 & 0.055 & 0.043 & 0.045 & 0.051 & 0.054 & 0.052\\
& A & 0.046 & 0.044 & 0.031 & 0.024 & 0.026 & 0.021 & 0.031 & 0.024\\
& Pear &0.064 & 0.038 & 0.040 & 0.034 & 0.022 & 0.010 & 0.016 & 0.013\\
\hline
\multirow{3}{*}{$\mathbf{Ga - M_2N}$} & P & 0.045 & 0.051 & 0.052 & 0.041 & 0.050 & 0.039 & 0.047 & 0.042\\
& A & 0.049 & 0.046 & 0.050 & 0.047 & 0.049 & 0.034 & 0.048 & 0.041\\
& Pear &0.052 & 0.047 & 0.062 & 0.052 & 0.063 & 0.049 & 0.051 & 0.049\\
\hline
\multirow{3}{*}{$\mathbf{Ga - M_3N}$} & P & 0.048 & 0.053 & 0.051 & 0.042 & 0.051 & 0.052 & 0.050 & 0.044\\
& A & 0.046 & 0.051 & 0.047 & 0.042 & 0.049 & 0.048 & 0.047 & 0.043\\
& Pear &0.035 & 0.043 & 0.051 & 0.052 & 0.039 & 0.040 & 0.044 & 0.047\\
\hline
\multirow{3}{*}{$\mathbf{Ga - M_2St}$} & P & 0.033 & 0.048 & 0.055 & 0.048 & 0.052 & 0.046 & 0.040 & 0.057\\
& A & 0.031 & 0.044 & 0.044 & 0.041 & 0.038 & 0.043 & 0.033 & 0.050\\
& Pear &0.049 & 0.049 & 0.047 & 0.052 & 0.051 & 0.050 & 0.036 & 0.049\\
\hline
\multirow{3}{*}{$\mathbf{SN - M_2N}$} & P & 0.052 & 0.053 & 0.040 & 0.053 & 0.042 & 0.055 & 0.050 & 0.046 \\
& A & 0.052 & 0.053 & 0.036 & 0.052 & 0.038 & 0.050 & 0.047 & 0.044 \\
& Pear &0.045 & 0.045 & 0.036 & 0.040 & 0.047 & 0.054 & 0.051 & 0.039\\
\hline
\multirow{3}{*}{$\mathbf{SN - M_3N}$} & P & 0.047 & 0.049 & 0.049 & 0.052 & 0.044 & 0.051 & 0.046 & 0.045 \\
& A & 0.045 & 0.045 & 0.048 & 0.048 & 0.045 & 0.050 & 0.040 & 0.045 \\
& Pear &0.039 & 0.048 & 0.067 & 0.052 & 0.045 & 0.061 & 0.052 & 0.051\\
\hline
\multirow{3}{*}{$\mathbf{SN - M_2St}$} & P & 0.050 & 0.043 & 0.058 & 0.044 & 0.055 & 0.056 & 0.056 & 0.056 \\
& A & 0.047 & 0.039 & 0.050 & 0.040 & 0.044 & 0.054 & 0.048 & 0.050 \\
& Pear &0.049 & 0.051 & 0.044 & 0.052 & 0.050 & 0.038 & 0.040 & 0.063\\
\hline
\multirow{3}{*}{$\mathbf{vM - M_2N}$} & P & 0.049 & 0.039 & 0.046 & 0.040 & 0.061 & 0.050 & 0.055 & 0.052 \\
& A & 0.054 & 0.035 & 0.042 & 0.036 & 0.057 & 0.040 & 0.047 & 0.046 \\
& Pear &0.053 & 0.046 & 0.048 & 0.049 & 0.052 & 0.050 & 0.055 & 0.059\\
\hline
\multirow{3}{*}{$\mathbf{vM - M_3N}$} & P & 0.049 & 0.056 & 0.042 & 0.051 & 0.049 & 0.048 & 0.044 & 0.053 \\
& A & 0.047 & 0.050 & 0.041 & 0.049 & 0.051 & 0.043 & 0.041 & 0.045 \\
& Pear &0.046 & 0.053 & 0.051 & 0.051 & 0.050 & 0.050 & 0.039 & 0.049\\
\hline
\multirow{3}{*}{$\mathbf{vM - M_2St}$} & P & 0.055 & 0.045 & 0.053 & 0.048 & 0.046 & 0.049 & 0.047 & 0.044 \\
& A & 0.052 & 0.038 & 0.044 & 0.042 & 0.037 & 0.044 & 0.041 & 0.034 \\
& Pear &0.051 & 0.044 & 0.045 & 0.046 & 0.042 & 0.043 & 0.046 & 0.036\\
\bottomrule
\bottomrule
\end{tabu}
\label{Simple Dcor}
\end{table}

\subsection{Case 2: Conditional independence I}
For the conditional independence case, we followed a similar experimental design. We generated three independent variables, $X$, $Y$, from the same sets of distributions used in the independence case, resulting in the same 15 unique distributional scenarios. The variable $Z$ was generated from a mixture of two components, namely exponential and the Weibull distributions. The empirical type I error rates are presented in Table \ref{cond1}. The permutation-based partial distance correlation test performed well in all cases, but its asymptotic version and the Pearson partial correlation test underestimated the nominal type I error in the case of Cauchy distribution and mixtures of two skew-t distributions ($\mathbf{Cau}-\mathbf{M_2St}$).

\begin{table}[h!]
\caption{Estimated type I error for Case 2: conditional independence I.}
\centering
\begin{tabu} to \linewidth {l|>{\raggedright}p{0.8cm}>{\raggedleft}p{0.8cm}>{\raggedleft}p{0.8cm}>{\raggedleft}p{0.8cm}>{\raggedleft}p{0.8cm}>{\raggedleft}p{0.8cm}>{\raggedleft}p{0.8cm}>{\raggedleft}p{0.8cm}>{\raggedleft}p{0.75cm}}
\toprule
\toprule
  \textbf{Distributions}& \textbf{N} & \textbf{50} & \textbf{100} & \textbf{200} & \textbf{500} & \textbf{1000} & \textbf{2000} & \textbf{5000} & \textbf{10000}\\
\midrule
\midrule
\multirow{3}{*}{$\mathbf{Be - M_2N}$} & P & 0.055 & 0.041 & 0.041 & 0.057 & 0.048 & 0.059 & 0.060 & 0.048\\
& A & 0.053 & 0.040 & 0.038 & 0.055 & 0.041 & 0.049 & 0.061 & 0.047\\
& PPear &0.038 & 0.043 & 0.042 & 0.040 & 0.040 & 0.041 & 0.044 & 0.045\\
\hline
\multirow{3}{*}{$\mathbf{Be - M_3N}$}& P & 0.051 & 0.049 & 0.042 & 0.052 & 0.056 & 0.043 & 0.050 & 0.037\\
& A & 0.052 & 0.049 & 0.038 & 0.049 & 0.046 & 0.040 & 0.049 & 0.036\\
& PPear &0.061 & 0.053 & 0.049 & 0.058 & 0.045 & 0.055 & 0.042 & 0.044\\
\hline
\multirow{3}{*}{$\mathbf{Be - M_2St}$} & P & 0.050 & 0.057 & 0.039 & 0.049 & 0.047 & 0.050 & 0.057 & 0.045\\
& A & 0.047 & 0.054 & 0.034 & 0.046 & 0.039 & 0.041 & 0.052 & 0.039\\
& PPear &0.040 & 0.038 & 0.051 & 0.046 & 0.045 & 0.044 & 0.045 & 0.042\\
\hline
\multirow{3}{*}{$\mathbf{Cau - M_2N}$} & P & 0.051 & 0.064 & 0.055 & 0.044 & 0.038 & 0.040 & 0.043 & 0.047\\
& A & 0.052 & 0.054 & 0.050 & 0.031 & 0.035 & 0.033 & 0.040 & 0.035\\
& PPear &0.026 & 0.042 & 0.046 & 0.058 & 0.042 & 0.062 & 0.055 & 0.045\\
\hline
\multirow{3}{*}{$\mathbf{Cau - M_3N}$} & P & 0.049 & 0.053 & 0.050 & 0.056 & 0.040 & 0.043 & 0.040 & 0.051\\
& A & 0.045 & 0.046 & 0.037 & 0.042 & 0.035 & 0.035 & 0.030 & 0.043\\
& PPear &0.046 & 0.053 & 0.053 & 0.052 & 0.053 & 0.044 & 0.049 & 0.062\\
\hline
\multirow{3}{*}{$\mathbf{Cau - M_2St}$} & P & 0.035 & 0.047 & 0.047 & 0.054 & 0.051 & 0.062 & 0.048 & 0.053\\
& A & 0.025 & 0.025 & 0.030 & 0.029 & 0.023 & 0.029 & 0.024 & 0.028\\
& PPear &0.055 & 0.037 & 0.028 & 0.031 & 0.031 & 0.014 & 0.011 & 0.006\\
\hline
\multirow{3}{*}{$\mathbf{Ga - M_2N}$} & P & 0.055 & 0.058 & 0.048 & 0.053 & 0.061 & 0.054 & 0.049 & 0.048\\
& A & 0.051 & 0.056 & 0.045 & 0.050 & 0.056 & 0.049 & 0.045 & 0.042\\
& PPear &0.042 & 0.039 & 0.066 & 0.049 & 0.035 & 0.045 & 0.052 & 0.033\\
\hline
\multirow{3}{*}{$\mathbf{Ga - M_3N}$} & P & 0.046 & 0.047 & 0.044 & 0.053 & 0.047 & 0.045 & 0.065 & 0.049\\
& A & 0.048 & 0.047 & 0.043 & 0.051 & 0.042 & 0.039 & 0.062 & 0.043\\
& PPear &0.038 & 0.056 & 0.045 & 0.054 & 0.048 & 0.031 & 0.046 & 0.061\\
\hline
\multirow{3}{*}{$\mathbf{Ga - M_2St}$} & P & 0.045 & 0.047 & 0.050 & 0.038 & 0.053 & 0.061 & 0.050 & 0.058\\
& A & 0.042 & 0.045 & 0.036 & 0.036 & 0.039 & 0.054 & 0.040 & 0.049\\
& PPear &0.049 & 0.047 & 0.042 & 0.037 & 0.046 & 0.039 & 0.044 & 0.052\\
\hline
\multirow{3}{*}{$\mathbf{SN - M_2N}$} & P & 0.047 & 0.052 & 0.045 & 0.054 & 0.045 & 0.049 & 0.057 & 0.056 \\
& A & 0.052 & 0.048 & 0.040 & 0.054 & 0.042 & 0.043 & 0.057 & 0.053 \\
& PPear &0.051 & 0.048 & 0.033 & 0.046 & 0.039 & 0.054 & 0.036 & 0.046\\
\hline
\multirow{3}{*}{$\mathbf{SN - M_3N}$} & P & 0.043 & 0.052 & 0.049 & 0.052 & 0.057 & 0.050 & 0.052 & 0.037 \\
& A & 0.040 & 0.050 & 0.042 & 0.045 & 0.056 & 0.046 & 0.048 & 0.039 \\
& PPear &0.043 & 0.049 & 0.044 & 0.053 & 0.048 & 0.067 & 0.051 & 0.044\\
\hline
\multirow{3}{*}{$\mathbf{SN - M_2St}$} & P & 0.058 & 0.047 & 0.041 & 0.039 & 0.060 & 0.059 & 0.052 & 0.053 \\
& A & 0.056 & 0.044 & 0.040 & 0.031 & 0.057 & 0.050 & 0.046 & 0.041 \\
& PPear &0.047 & 0.054 & 0.048 & 0.048 & 0.054 & 0.049 & 0.039 & 0.040\\
\hline
\multirow{3}{*}{$\mathbf{vM - M_2N}$} & P & 0.051 & 0.048 & 0.051 & 0.046 & 0.060 & 0.049 & 0.037 & 0.042 \\
& A & 0.049 & 0.044 & 0.046 & 0.043 & 0.053 & 0.048 & 0.037 & 0.040\\
& PPear &0.049 & 0.042 & 0.048 & 0.055 & 0.047 & 0.050 & 0.058 & 0.059\\
\hline
\multirow{3}{*}{$\mathbf{vM - M_3N}$} & P & 0.064 & 0.042 & 0.051 & 0.045 & 0.052 & 0.046 & 0.055 & 0.053 \\
& A & 0.064 & 0.037 & 0.050 & 0.043 & 0.048 & 0.039 & 0.049 & 0.049 \\
& PPear &0.063 & 0.063 & 0.043 & 0.055 & 0.050 & 0.051 & 0.039 & 0.051\\
\hline
\multirow{3}{*}{$\mathbf{vM - M_2St}$} & P & 0.050 & 0.042 & 0.046 & 0.062 & 0.051 & 0.061 & 0.056 & 0.048 \\
& A & 0.049 & 0.035 & 0.041 & 0.048 & 0.042 & 0.053 & 0.046 & 0.043 \\
& PPear &0.035 & 0.047 & 0.035 & 0.055 & 0.056 & 0.045 & 0.041 & 0.048\\
\bottomrule
\bottomrule
\end{tabu}
\label{cond1}
\end{table}

\subsection{Case 3: Conditional independence IΙ}
For the conditional independence case, we followed a similar experimental design. We generated three independent variables, $X$, $Y$, from the same sets of distributions used in the independence case, resulting in the same 15 unique distributional scenarios and again, but this time, the variable $Z$, generated again from the mixture of exponential-Weibull distributions and was added in $X$ and $Y$. The empirical type I error rates are presented in Table \ref{cond2}. The partial distance correlation test failed in call cases, regardless of the approach used to compute the p-value. Surprisingly enough the Pearson correlation succeeded except for one case, that of Cauchy distribution and mixtures of two skew-t distributions ($\mathbf{Cau}-\mathbf{M_2St}$).

\begin{table}[h!]
\caption{Estimated type I error for Case 3: conditional independence IΙ.}
\centering
\begin{tabu} to \linewidth {l|>{\raggedright}p{0.8cm}>{\raggedleft}p{0.8cm}>{\raggedleft}p{0.8cm}>{\raggedleft}p{0.8cm}>{\raggedleft}p{0.8cm}>{\raggedleft}p{0.8cm}>{\raggedleft}p{0.8cm}>{\raggedleft}p{0.8cm}>{\raggedleft}p{0.75cm}}
\toprule
\toprule
  \textbf{Distributions}& \textbf{N} & \textbf{50} & \textbf{100} & \textbf{200} & \textbf{500} & \textbf{1000} & \textbf{2000} & \textbf{5000} & \textbf{10000}\\
\midrule
\midrule
\multirow{3}{*}{$\mathbf{Be - M_2N}$} & P & 0.200 & 0.256 & 0.345 & 0.394 & 0.487 & 0.513 & 0.575 & 0.570\\
& A & 0.195 & 0.256 & 0.347 & 0.390 & 0.486 & 0.513 & 0.575 & 0.571\\
& Pear &0.054 & 0.038 & 0.054 & 0.048 & 0.063 & 0.040 & 0.053 & 0.041\\
\hline
\multirow{3}{*}{$\mathbf{Be - M_3N}$} & P & 0.172 & 0.262 & 0.335 & 0.423 & 0.485 & 0.512 & 0.564 & 0.560\\
& A & 0.171 & 0.258 & 0.331 & 0.421 & 0.484 & 0.514 & 0.566 & 0.559\\
& Pear &0.052 & 0.059 & 0.043 & 0.041 & 0.059 & 0.047 & 0.049 & 0.054
\\
\hline
\multirow{3}{*}{$\mathbf{Be - M_2St}$} & P & 0.189 & 0.208 & 0.268 & 0.330 & 0.375 & 0.447 & 0.540 & 0.583\\
& A & 0.178 & 0.190 & 0.255 & 0.315 & 0.359 & 0.440 & 0535 & 0.577\\
& Pear &0.048 & 0.037 & 0.042 & 0.045 & 0.043 & 0.052 & 0.054 & 0.049\\
\hline
\multirow{3}{*}{$\mathbf{Cau - M_2N}$} & P & 0.079 & 0.076 & 0.091 & 0.131 & 0.171 & 0.230 & 0.290 & 0.377\\
& A & 0.077 & 0.063 & 0.083 & 0.113 & 0.163 & 0.211 & 0.278 & 0.365\\
& Pear &0.059 & 0.046 & 0.044& 0.054 & 0.046 & 0.047 & 0.050 & 0.054\\
\hline
\multirow{3}{*}{$\mathbf{Cau - M_3N}$} & P & 0.064 & 0.074 & 0.078 & 0.127 & 0.157 & 0.226 & 0.293 & 0.355\\
& A & 0.060 & 0.064 & 0.065 & 0.117 & 0.138 & 0.208 & 0.283 & 0.344\\
& Pear &0.048 & 0.058 & 0.055 & 0.058 & 0.051 & 0.046 & 0.051 & 0.043\\
\hline
\multirow{3}{*}{$\mathbf{Cau - M_2St}$} & P & 0.073 & 0.076 & 0.085 & 0.088 & 0.118 & 0.156 & 0.229 & 0.291\\
& A & 0.062 & 0.048 & 0.058 & 0.061 & 0.081 & 0.091 & 0.161 & 0.209\\
& Pear &0.045 & 0.039 & 0.044 & 0.029 & 0.022 & 0.015 & 0.011 & 0.011\\
\hline
\multirow{3}{*}{$\mathbf{Ga - M_2N}$} & P & 0.201 & 0.246 & 0.322 & 0.413 & 0.487 & 0.539 & 0.591 & 0.631\\
& A & 0.198 & 0.246 & 0.322 & 0.405 & 0.489 & 0.537 & 0.590 & 0.630\\
& Pear &0.050 & 0.048 & 0.051 & 0.047 & 0.052 & 0.038 & 0.060 & 0.050\\
\hline
\multirow{3}{*}{$\mathbf{Ga - M_3N}$} & P & 0.200 & 0.253 & 0.337 & 0.412 & 0.473 & 0.518 & 0.578 & 0.629\\
& A & 0.194 & 0.249 & 0.337 & 0.414 & 0.470 & 0.519 & 0.578 & 0.629\\
& Pear &0.048 & 0.047 & 0.050 & 0.044 & 0.051 & 0.058 & 0.053 & 0.057\\
\hline
\multirow{3}{*}{$\mathbf{Ga - M_2St}$} & P & 0.190 & 0.253 & 0.268 & 0.359 & 0.418 & 0.492 & 0.619 & 0.647\\
& A & 0.180 & 0.240 & 0.260 & 0.349 & 0.415 & 0.482 & 0.610 & 0.647\\
& Pear &0.038 & 0.055 & 0.052 & 0.043 & 0.042 & 0.048 & 0.047 & 0.042\\
\hline
\multirow{3}{*}{$\mathbf{SN - M_2N}$} & P & 0.161 & 0.230 & 0.271 & 0.403 & 0.432 & 0.479 & 0.536 & 0.590 \\
& A & 0.162 & 0.219 & 0.268 & 0.395 & 0.424 & 0.479 & 0.532 & 0.586 \\
& Pear &0.052 & 0.047 & 0.041 & 0.052 & 0.046 & 0.057 & 0.052 & 0.045\\
\hline
\multirow{3}{*}{$\mathbf{SN - M_3N}$} & P & 0.192 & 0.233 & 0.273 & 0.375 & 0.467 & 0.496 & 0.555 & 0.568 \\
& A & 0.190 & 0.225 & 0.269 & 0.369 & 0.460 & 0.494 & 0.553 & 0.568 \\
& Pear &0.050 & 0.048 & 0.049 & 0.048 & 0.044 & 0.063 & 0.042 & 0.039\\
\hline
\multirow{3}{*}{$\mathbf{SN - M_2St}$} & P & 0.159 & 0.169 & 0.246 & 0.315 & 0.357 & 0.463 & 0.521 & 0.560 \\
& A & 0.146 & 0.166 & 0.229& 0.308 & 0.349 & 0.454 & 0.514 & 0.553 \\
& Pear &0.041 & 0.054 & 0.051 & 0.040 & 0.048 & 0.060 & 0.049 & 0.062\\
\hline
\multirow{3}{*}{$\mathbf{vM - M_2N}$} & P & 0.188 & 0.227 & 0.306 & 0.359 & 0.432 & 0.497 & 0.584 & 0.607 \\
& A & 0.182 & 0.216 & 0.308 & 0.361 & 0.428 & 0.494 & 0.581 & 0.604 \\
& Pear &0.049 & 0.035 & 0.042 & 0.043 & 0.056 & 0.048 & 0.046 & 0.046\\
\hline
\multirow{3}{*}{$\mathbf{vM - M_3N}$} & P & 0.141 & 0.234 & 0.305 & 0.418 & 0.471 & 0.505 & 0.604 & 0.630 \\
& A & 0.143 & 0.233 & 0.296 & 0.412 & 0.467 & 0.503 & 0.604 & 0.630 \\
& Pear &0.036 & 0.057 & 0.073 & 0.058 & 0.061 & 0.033 & 0.056 & 0.053\\
\hline
\multirow{3}{*}{$\mathbf{vM - M_2St}$} & P & 0.198 & 0.217 & 0.255 & 0.354 & 0.361 & 0.458 & 0.563 & 0.634 \\
& A & 0.184 & 0.201 & 0.232 & 0.338 & 0.351 & 0.450 & 0.555 & 0.626 \\
& Pear &0.052 & 0.047 & 0.057 & 0.043 & 0.041 & 0.051 & 0.054 & 0.044\\
\bottomrule
\bottomrule
\end{tabu}
\label{cond2}
\end{table}

\subsection{Case 4: Non-linear conditional independence}
In contrast to case 3, where $X$, $Y$ and $Z$ were independent, this time a conditional association is introduced. According to the theory of Bayesian networks \citep{neapolitan2003}, when $Z$ affects $X$ and $Z$ affects $Y$, independently, the two variables $X$ and $Y$ are dependent, but conditional on $Z$ they are independent. Figure \ref{dag}(a) shows this graphically. The values of $Z$ were generated from the 8 aforementioned distributions and the values of $X$ and $Y$ were generated as follows:
\begin{eqnarray*}
x_i &=& \log{\left(|z_i|\right)} + z_i^2 + \eta_i \\
y_i &=& \sin{\left(z_i\right)} + \log_{10}{\left(|z_i|\right)} + \eta_i,
\end{eqnarray*}
where $\eta_i \sim N(0,1)$.

The estimated type I error rates appear on Table \ref{cond3}. This example clearly shows the failure of the Pearson partial correlation to assess conditional independence, but it also demonstrates the failure of the partial distance correlation tests, which is surprisingly enough. 

\begin{table}[h!]
\caption{Estimated type I error for Case 4: non-linear conditional independence.}
\centering
\begin{tabu} to \linewidth {c|>{\raggedright}p{0.8cm}>{\raggedleft}p{0.8cm}>{\raggedleft}p{0.8cm}>{\raggedleft}p{0.8cm}>{\raggedleft}p{0.8cm}>{\raggedleft}p{0.8cm}>{\raggedleft}p{0.8cm}>{\raggedleft}p{0.8cm}>{\raggedleft}p{0.75cm}}
\toprule
\toprule
  \textbf{Distributions}& \textbf{N} & \textbf{50} & \textbf{100} & \textbf{200} & \textbf{500} & \textbf{1000} & \textbf{2000} & \textbf{5000} & \textbf{10000}\\
\midrule
\midrule
\multirow{3}{*}{$\mathbf{Be}$} & P & 0.085 & 0.115 &0.146 & 0.175 &0.239 &0.271 &0.345 &0.413 \\
& A & 0.089 &0.110 &0.142 &0.166 &0.235 &0.273 &0.335 &0.410 \\
& Pear & 0.052 &0.052 &0.051 &0.074 &0.100 &0.139 &0.174 &0.169 \\
\hline
\multirow{3}{*}{$\mathbf{Cau}$} & P & 0.316 & 0.421 & 0.483 & 0.471 & 0.379 & 0.212 & 0.071 & 0.019 \\
& A &0.305 & 0.411 & 0.471 &0.458 & 0.365 & 0.205 & 0.069 & 0.019 \\
& Pear & 0.201 & 0.309 & 0.417 &0.509 & 0.579 & 0.634& 0.706& 0.716 \\
\hline
\multirow{3}{*}{$\mathbf{Ga}$} & P &0.265 &0.341 &0.436 &0.535 &0.581 &0.676 &0.729 &0.808 \\
& A &0.266 &0.345 &0.432 &0.528 &0.577 &0.676 &0.727 &0.805 \\
& Pear &0.178 &0.191 &0.192 &0.195 &0.214 &0.252 &0.377 &0.511 \\
\hline
\multirow{3}{*}{$\mathbf{SN}$} & P &0.115 &0.169 &0.186 &0.173 &0.183 &0.172 &0.146 &0.139 \\
& A &0.122 &0.167 &0.184 &0.172 &0.182 &0.172 &0.146 &0.139 \\
& Pear &0.097 &0.171 &0.232 &0.361 &0.464 &0.599 &0.801 &0.903 \\
\hline
\multirow{3}{*}{$\mathbf{vM}$} & P &0.052 &0.112 &0.177 &0.285 &0.346 &0.443 &0.478 &0.517 \\
& A & 0.052&0.108 &0.169 &0.280 &0.343 &0.441 &0.476 &0.517 \\
& Pear &0.144 &0.207 &0.316 &0.460 &0.545 &0.627 &0.659 &0.721 \\
\hline
\multirow{3}{*}{$\mathbf{M_2N}$} & P & 0.219 &0.400 & 0.551 &0.658 &0.679 & 0.711&0.690 &0.715 \\
& A & 0.216&0.390 &0.553 &0.657 &0.674 & 0.708 & 0.690 & 0.715 \\
& Pear &0.157 &0.291 &0.417 & 0.593 &0.691 & 0.753 & 0.757 & 0.776 \\
\hline
\multirow{3}{*}{$\mathbf{M_3N}$} & P & 0.221 &0.384 & 0.529 & 0.666 & 0.690 & 0.692 & 0.711 & 0.714 \\
& A & 0.222&0.379 &0.526 &0.664 & 0.687&0.692 &0.710 & 0.714 \\
& Pear & 0.161& 0.261&0.395 &0.606 & 0.706 & 0.738 & 0.775 & 0.751 \\
\hline
\multirow{3}{*}{$\mathbf{M_2St}$} & P & 0.096 & 0.109 & 0.108 & 0.086 & 0.049 & 0.012 & 0.006 &0.002 \\
& A & 0.090 & 0.106 & 0.103 & 0.082 & 0.045 & 0.011 & 0.006 & 0.001 \\
& Pear & 0.114 & 0.166 &0.304 & 0.544 & 0.736 & 0.870 & 0.931&0.967 \\
\bottomrule
\bottomrule
\end{tabu}
\label{cond3}
\end{table}

\subsection{Case 5: Conditional dependence}
\begin{figure}[!ht]
\centering
\begin{tabular}{cc}
\includegraphics[scale = 0.4]{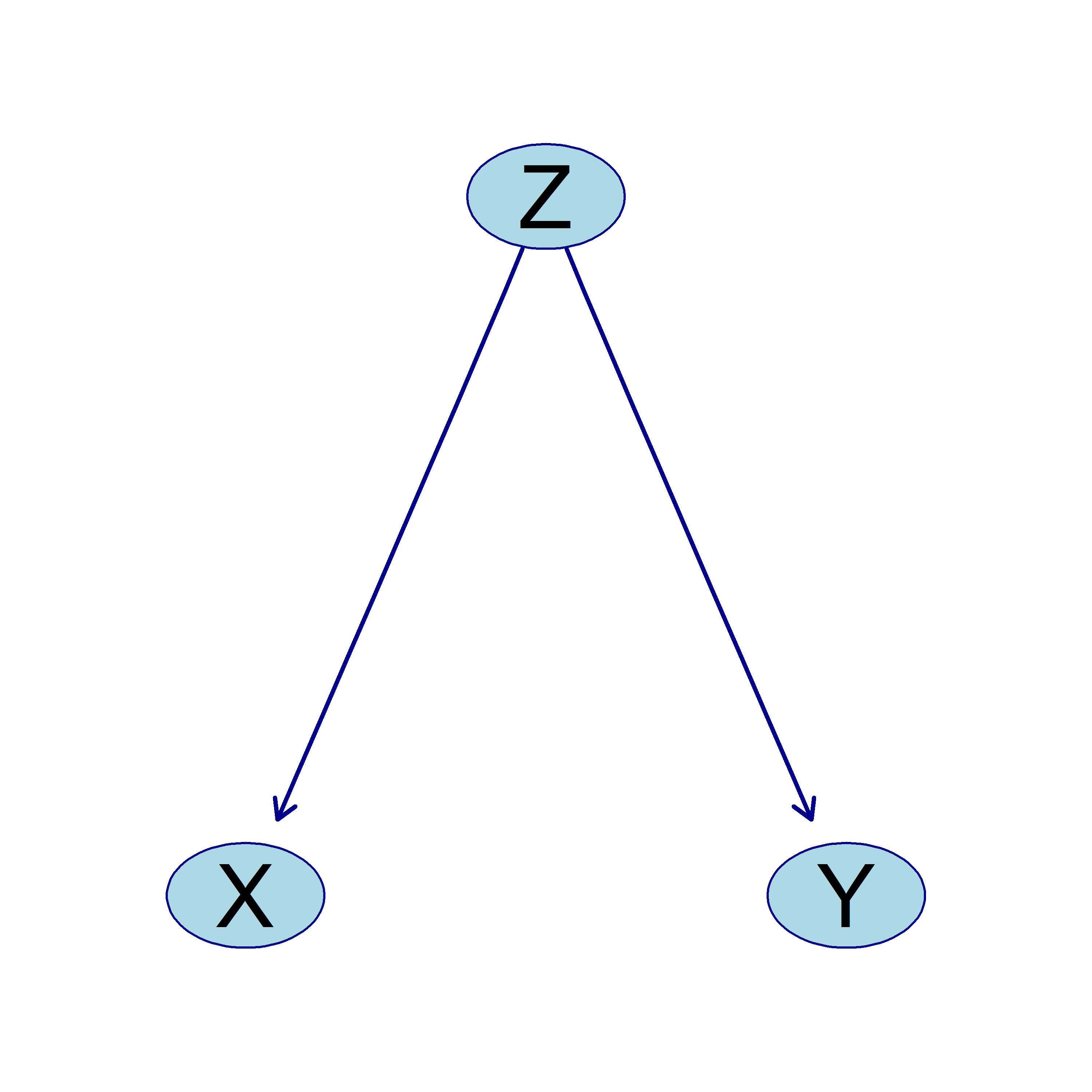} &
\includegraphics[scale = 0.4]{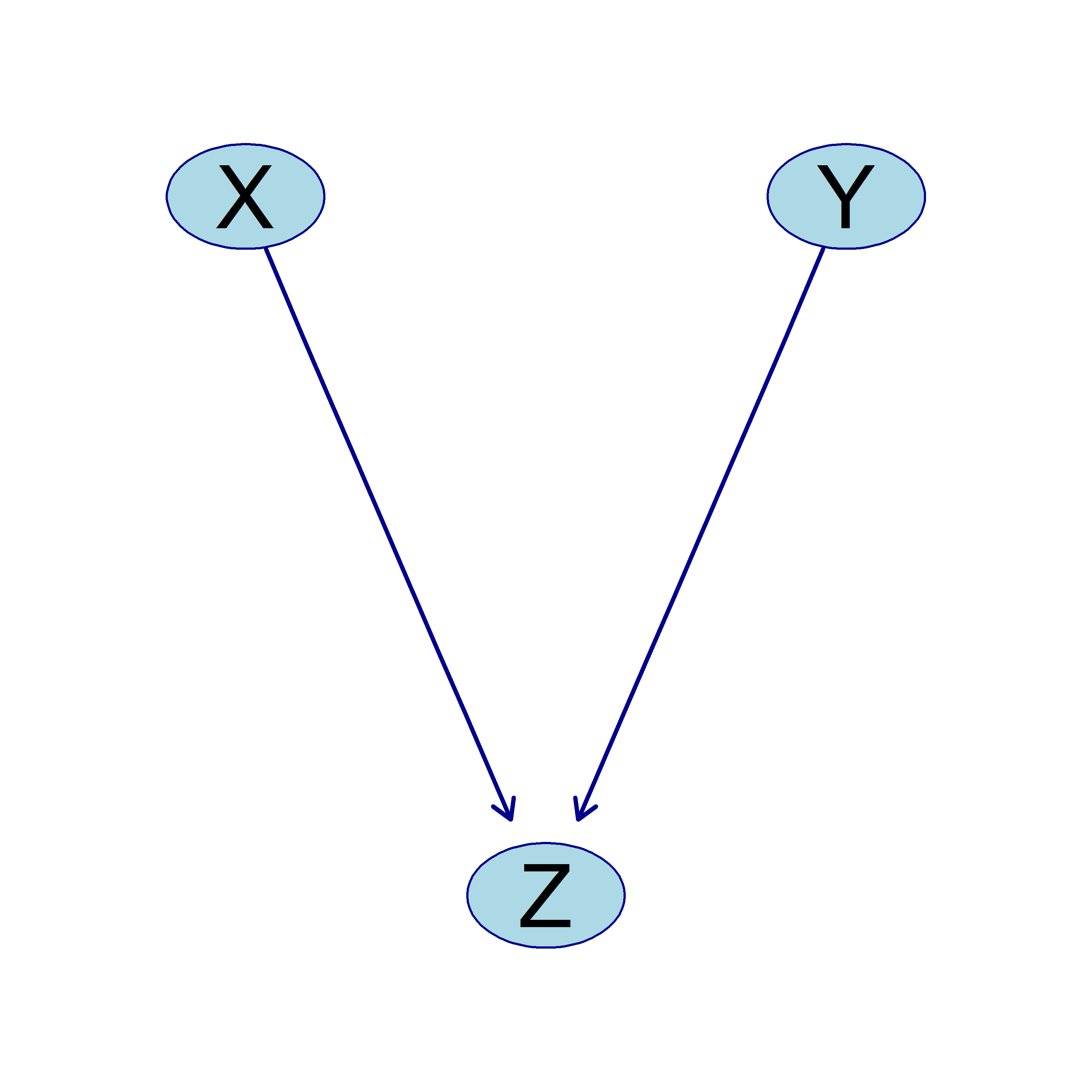} \\
(a) Conditional independence ($X \CI Y \vert Z$).  &  (b) Conditional dependence ($X \nCI Y \vert Z$).
\end{tabular}
\caption{DAG visualizations of the conditional independence and conditional dependence cases.}
\label{dag}
\end{figure}

For this case, the opposite case is examined. Following the theory of Bayesian networks \citep{neapolitan2003} again, when two independent variables $X$ and $Y$ jointly affect $Z$, they become dependent conditioning on $Z$. Figure \ref{dag}(b) visualizes the conditional dependence relationship. The values of $X$ and $Y$ were generated in the same manner as previously (15 scenarios), but the values of $Z$ and were generated as follows
\begin{eqnarray*}
z_i &=& \log{\left(|x_i|\right)} + \sin{\left(y_i\right)} + \eta_i \\
\end{eqnarray*}
where $\eta_i \sim N(0,1)$.

In the simulations conducted by \cite{zhu2017}, there were cases where the unconditional distance correlation test had no power at all. Our example corroborates those results, as Table \ref{conddep} shows that the estimated power of the partial conditional correlation never exceeds $6\%$.

\begin{table}[h!]
\caption{Estimated power for Case 4: conditional dependence.}
\centering
\begin{tabu} to \linewidth {l|>{\raggedright}p{0.8cm}>{\raggedleft}p{0.8cm}>{\raggedleft}p{0.8cm}>{\raggedleft}p{0.8cm}>{\raggedleft}p{0.8cm}>{\raggedleft}p{0.8cm}>{\raggedleft}p{0.8cm}>{\raggedleft}p{0.8cm}>{\raggedleft}p{0.8cm}>{\raggedleft}p{0.8cm}>{\raggedleft}p{0.8cm}}
\toprule
\toprule
\textbf{Distributions} & \textbf{n} & \textbf{50} & \textbf{100} & \textbf{150} & \textbf{200} & \textbf{250} & \textbf{300} & \textbf{350} & \textbf{400} & \textbf{450} & \textbf{500} \\
\midrule
\midrule
\multirow{3}{*}{$\mathbf{Be - M_2N}$} & P & 0.054 & 0.042 & 0.041 & 0.044 & 0.054 & 0.038 & 0.052& 0.055 & 0.040& 0.035\\
& A & 0.058& 0.037& 0.039 & 0.040 & 0.049 & 0.037 & 0.050& 0.053 & 0.038& 0.033\\
& Pear & 0.058& 0.049 & 0.067& 0.072& 0.076 & 0.075 & 0.091 & 0.088 & 0.092& 0.090\\
\hline
\multirow{3}{*}{$\mathbf{Be - M_3N}$} & P &0.043 & 0.040& 0.064 &0.037 & 0.041& 0.053& 0.042 & 0.039 & 0.052&0.045 \\
& A &0.041 & 0.040& 0.051& 0.036 & 0.039& 0.047 & 0.043& 0.035& 0.049 &0.040 \\
& Pear &0.049 & 0.053& 0.071 & 0.061 & 0.070 & 0.096 & 0.093&0.086 & 0.105& 0.090\\
\hline
\multirow{3}{*}{$\mathbf{Be - M_2St}$} & P & 0.057 & 0.041& 0.048 & 0.047& 0.049& 0.042& 0.048 & 0.046& 0.043& 0.040\\
& A & 0.054& 0.038 & 0.045& 0.042 & 0.042 & 0.039 & 0.045 & 0.040& 0.038& 0.036 \\
& Pear &0.044 & 0.044 & 0.066 & 0.050&0.048 & 0.061 & 0.043 & 0.057 & 0.061&0.053 \\
\hline
\multirow{3}{*}{$\mathbf{Cau - M_2N}$} & P & 0.048& 0.062& 0.051&0.048 & 0.040& 0.059& 0.046&0.045 & 0.045&0.049 \\
& A &0.044 &0.057 &0.046 &0.043 & 0.034 & 0.048& 0.039&0.037 &0.041 &0.045 \\
& Pear & 0.038 & 0.060 & 0.064& 0.056& 0.045 & 0.056& 0.047& 0.061& 0.061&0.061 \\
\hline
\multirow{3}{*}{$\mathbf{Cau - M_3N}$} & P &0.045 &0.052 & 0.043 & 0.048& 0.044& 0.050& 0.045 & 0.059& 0.038&0.053 \\
& A &0.043 &0.047 & 0.038 & 0.041 & 0.035& 0.040& 0.042&0.054 &0.035 & 0.044\\
& Pear & 0.047& 0.039 &0.045 & 0.046&0.061 & 0.051&0.049 & 0.056& 0.055& 0.058\\
\hline
\multirow{3}{*}{$\mathbf{Cau - M_2St}$} & P & 0.049 &0.043 &0.042 & 0.049&0.046 & 0.046& 0.048 & 0.055 & 0.049 &0.040 \\
& A &0.033 &0.023 &0.025 &0.031 &0.032 &0.028 & 0.026 &0.027 & 0.030&0.023 \\
& Pear &0.047 &0.047 &0.033 & 0.038& 0.037&0.026 & 0.024&0.026 & 0.038&0.029 \\
\hline
\multirow{3}{*}{$\mathbf{Ga - M_2N}$} & P & 0.059& 0.041& 0.048 & 0.042 & 0.035& 0.046 & 0.046 & 0.049 & 0.042 & 0.050\\
& A &0.059 & 0.037 & 0.044 & 0.043 & 0.035& 0.041& 0.042 & 0.041 & 0.042&0.047 \\
& Pear & 0.055&0.040 & 0.054 & 0.046& 0.053&0.042 & 0.050 & 0.061 & 0.040& 0.050\\
\hline
\multirow{3}{*}{$\mathbf{Ga - M_3N}$} & P & 0.048& 0.053 & 0.041 & 0.059 &0.040 &0.048 & 0.054& 0.043&0.054 &0.058 \\
& A &0.044 & 0.043 & 0.038& 0.055 &0.038 & 0.044& 0.050 &0.039 &0.049 &0.049 \\
& Pear &0.040 & 0.047 & 0.045 & 0.060 &0.044 &0.049 & 0.051 &0.043 &0.053 &0.048 \\
\hline
\multirow{3}{*}{$\mathbf{Ga - M_2St}$} & P & 0.056& 0.040& 0.047 & 0.034 &0.060 & 0.049 & 0.032& 0.036 & 0.039 & 0.033 \\
& A & 0.050 & 0.033&0.042 & 0.027 & 0.053 & 0.043 & 0.024 & 0.034 &0.033 & 0.029 \\
& Pear &0.041 & 0.040 & 0.051 & 0.043 & 0.056 & 0.061 & 0.043 & 0.053& 0.050& 0.039 \\
\hline
\multirow{3}{*}{$\mathbf{SN - M_2N}$} & P & 0.048& 0.046 & 0.060& 0.050&0.059 & 0.056 & 0.048 & 0.046  & 0.055 & 0.054 \\
& A & 0.043&0.042 &0.059 & 0.046& 0.055 & 0.046 & 0.046 & 0.041 & 0.047 & 0.049\\
& Pear & 0.043 & 0.049&0.057 &0.039 &0.052 & 0.047 & 0.055 &0.051 & 0.066 &0.054 \\
\hline
\multirow{3}{*}{$\mathbf{SN - M_3N}$} & P & 0.045 & 0.050 & 0.041 & 0.055&0.049 & 0.041 &0.075 &0.040 & 0.053 & 0.035 \\
& A & 0.047& 0.050&0.044 & 0.50 &0.047 &0.036 &0.068 &0.040 & 0.051 & 0.031\\
& Pear & 0.038& 0.048& 0.049 &0.053 &0.047 &0.039 &0.058 &0.045 & 0.048 & 0.043\\
\hline
\multirow{3}{*}{$\mathbf{SN - M_2St}$} & P & 0.055 & 0.045 & 0.046 & 0.049&0.053 & 0.056 &0.053 &0.045 & 0.041 & 0.055 \\
& A & 0.053& 0.040&0.043 & 0.43 &0.047 &0.048 &0.044 &0.036 & 0.038 & 0.054\\
& Pear & 0.053& 0.041& 0.042 &0.042 &0.054 &0.051 &0.060 &0.058 & 0.049 & 0.057\\
\hline
\multirow{3}{*}{$\mathbf{vM - M_2N}$} & P & 0.045 & 0.048 & 0.056 & 0.046 &0.052 & 0.054 & 0.045 & 0.042& 0.052& 0.042 \\
& A &0.043 & 0.046 & 0.053 & 0.043 & 0.048 & 0.053 & 0.044 & 0.034&0.051 &0.038 \\
& Pear & 0.051 & 0.053 & 0.062 & 0.070 & 0.086 & 0.010 & 0.010& 0.095& 0.094 & 0.112\\
\hline
\multirow{3}{*}{$\mathbf{vM - M_3N}$} & P & 0.041 & 0.048 & 0.040&0.057 & 0.036& 0.059 & 0.043& 0.042 & 0.045 & 0.054 \\
& A & 0.044 & 0.043 &0.038 & 0.052&0.033 & 0.056 & 0.042 & 0.040 & 0.043 & 0.046\\
& Pear & 0.043 &0.055 &0.074 &0.091 & 0.076 & 0.084 & 0.072 & 0.101 & 0.092 & 0.102 \\
\hline
\multirow{3}{*}{$\mathbf{vM - M_2St}$} & P & 0.037 &0.056 &0.061 &0.059 & 0.047& 0.047& 0.038 & 0.049 & 0.035 &0.043 \\
& A & 0.034 &0.048 &0.054 &0.056 & 0.046 & 0.039 & 0.033 & 0.042 & 0.030 & 0.036\\
& Pear & 0.043 &0.034 &0.043 &0.047 &0.059 & 0.052 & 0.055&0.059 & 0.046 &0.051 \\
\bottomrule
\bottomrule
\end{tabu}
\label{conddep}
\end{table}

\subsection{Case 6: Univariate filtering}
Following \cite{han2022} we reproduced an example where we use the distance and the Pearson correlation to detect the significant variables. We employed 8 of the aforementioned distributions, namely the Beta, Skew normal, von-Mises, Gamma, Cauchy, mixtures of two normals and mixtures of three normals and generated 50 variables from each distribution. This resulted in a total of 400 predictors. The response variable was generated as $Y_i=e^{{\bf X}_i^\top{b}+5}+\eta_i$, $i=1,\ldots,n$, where $\eta \sim N(0,1)$, $\bf X$ is the matrix with the standardized data and ${\bf b}=\left({\bf 1}_{10}, {\bf 0}_{790}\right)^\top$ is the vector of coefficients containing 10 non-zero coefficients. The sample sizes considered were $n=(100, 200, 300, 500, 1000)$.

According to Table \ref{filter} the distance correlation using the permutation-based p-value selected, on average, the highest number of truly statistically significant variables (P$_T$) but it also had the highest falsely selected number of variables (P$_F$). The distance correlation using the asymptotic p-value performed worse in identifying the statistically significant variables but it selected less irrelevant variables. Pearson correlation on the other hand selected less significant variables and many irrelevant variables. Finally, the last three columns of Table \ref{filter} show a high number of commonly selected variables by the distance correlation using either the permutation-based or the asymptotic p-value, but a small number of common variables identified by Pearson correlation.

\begin{table}[ht]
\caption{Results for Case 6: univariate fitering. The subsrcipts $T$ and $F$ indicate the number of significant variables identified and the number of falsely selected variables, respectively. The symbol $\cap$ denotes the number of commonly selected variables by pairs of the two approaches.}
\centering
\begin{tabular}{l|rr|rr|rr|rrr}
\bottomrule
Sample size & P$_T$ & P$_F$ & A$_T$ & A$_F$ & Pear$_T$ & Pear$_F$ & P$\cap$A & P$\cap$Pear & A$\cap$Pear \\ 
 \midrule
100   & 3.7 & 19.9 & 3.6 & 17.2 & 2.7 & 18.0 & 20.1 & 7.8 & 7.6 \\ 
200   & 4.2 & 19.8 & 4.0 & 16.2 & 2.7 & 17.7 & 19.6 & 7.4 & 7.1 \\ 
300   & 4.2 & 19.6 & 4.0 & 15.1 & 2.7 & 17.4 & 18.6 & 7.3 & 7.0 \\ 
500   & 4.0 & 29.6 & 3.4 & 13.7 & 2.5 & 17.4 & 16.6 & 7.3 & 6.3 \\ 
1000  & 4.3 & 45.7 & 3.2 & 12.1 & 2.6 & 17.0 & 14.9 & 7.7 & 5.8 \\ 
   \hline
\end{tabular}
\label{filter}
\end{table}

\subsection{Computational cost}
Regarding the computational cost to calculate the distance correlation between two pairs of univariate variables, \cite{Huo2016} proposed a very fast and memory efficient algorithm. However, when it comes to performing 499 permutations, the execution time increases. Table \ref{tab:cost} shows the time (in seconds) required by the permutation-based and the asymptotic (partial) distance correlation tests. When the sample size is equal to 50,000, the permutation-based distance correlation test is nearly 70 times slower than the asymptotic. But, when it comes to the partial distance correlation test, the permutation-based test is nearly 300 times slower than the asymptotic.  

\begin{table}[h]
\caption{Computational cost of (in seconds) of the (partial) distance correlation tests. The first two rows refer to the permutation-based test and are based on 499 permutations. The next two rows refer to the asymptotic test and report the time required to compute a single (partial) distance correlation.}
\centering
\begin{tabular}{l|cccccccccc}
\toprule
                   & \multicolumn{10}{c}{Sample size} \\ \midrule
\textit{R} function  & 50 & 100 & 200 & 500 & 1,000 & 2,000 & 5,000 & 10,000 & 20,000 & 50,000 \\
\midrule
\texttt{pdcor.test()} & 0.02 & 0.02 & 0.04 & 0.08 & 0.16 & 0.33 & 0.91 & 1.9 & 5.47 &  14.83 \\
\hline
\texttt{dcor.test()}  & 0.002 & 0.006 & 0.01 & 0.03 & 0.08 & 0.18 & 0.53 & 1.13 & 3.24 & 8.88 \\
\midrule
\texttt{pdcor()} & 0.00004 & 0.00004 &  0.00006 & 0.0001 & 0.0003 & 0.001 & 0.003 & 0.007 & 0.019 & 0.05 \\
\midrule
\texttt{dcor()} & 0.00002 & 0.00002 & 0.00002 & 0.00006 & 0.0001 & 0.0003 & 0.0009 & 0.001 & 0.005 & 0.013 \\
\bottomrule
\end{tabular}
\label{tab:cost}
\end{table}

\clearpage
\section{Conclusions} \label{sec:conclusions}
The simulation studies demonstrated multiple facets of the (conditional) independence testing. When it comes to simple independence between $X$ and $Y$ or between $X$ and $Y$ conditional on some $Z$, when there is no relationship among the variables, then regardless of the distribution of each variable, the distance correlation and the Pearson correlation work equally well in terms of type I error. However, if the same noise is added in $X$ and $Y$, then the partial distance correlation fails to identify the independence, whereas, surprisingly enough, Pearson correlation succeeds.

However, when non-linear dependence exists between $Z$ and $Y$ and between $Z$ and $X$, in such a way that $Z$ affects both $X$ and $Y$, then the independence of $X$ and $Y$ conditional on $Z$ is not detected by either correlation. On the antipode, when both $X$ and $Y$ affect $Z$, the conditional dependence is again not detected by either correlation. Finally, the distance correlation performed poorly for univariate filtering purposes. 

This paper demonstrated some of the weaknesses of the distance correlation that were also presented in \cite{zhu2017} who proposed the projection distance correlation as a better alternative to the distance correlation. However, a partial version of this newer correlation does not exist, neither a memory efficient algorithm to compute it fast. The final conclusion of this paper though is that perhaps we should start moving beyond the distance correlation. 

\bibliographystyle{chicago}
\bibliography{vivlio}

\end{document}